\documentclass{article}

\usepackage[preprint]{neurips_2024}
\setcitestyle{numbers}
\usepackage{graphicx}
\usepackage{amsmath}
\usepackage{amssymb}
\usepackage{listings}
\usepackage{hyperref}
\usepackage{xcolor}
\usepackage{booktabs}
\usepackage{multirow}
\usepackage{float}
\usepackage{caption}
\usepackage{listings}
\usepackage[ruled,vlined]{algorithm2e}
\SetKwProg{Fn}{Function}{}{}
\SetKwInOut{KwIn}{Input}
\SetKwInOut{KwOut}{Output}

\title{MLCPD: A Unified Multi-Language Code Parsing Dataset with Universal AST Schema}

\author{
  Jugal Gajjar\thanks{Corresponding author: \texttt{812jugalgajjar@gmail.com}} \\
  Department of Computer Science\\
  The George Washington University\\
  Washington, D.C. 20052 \\
  \texttt{jugal.gajjar@gwu.edu} \\
  \And
  Kamalasankari Subramaniakuppusamy \\
  Department of Computer Science\\
  The George Washington University\\
  Washington, D.C. 20052 \\
  \texttt{kamalasankaris@gwu.edu} \\
}

\begin{document}
\maketitle

\begin{abstract}
We introduce the MultiLang Code Parser Dataset (MLCPD), a large-scale, language-agnostic dataset unifying syntactic and structural representations of code across ten major programming languages. MLCPD contains over seven million parsed source files normalized under our proposed \emph{universal Abstract Syntax Tree (AST)} schema, enabling consistent cross-language reasoning, structural learning, and multilingual software analysis. Unlike existing corpora that focus purely on token-level code or isolated parsers, MLCPD provides both hierarchical tree representations and rich metadata for every file, ensuring lossless syntactic coverage and structural uniformity. Each entry includes a normalized schema, language-level metadata, and abstracted node semantics stored in Parquet format for scalable retrieval. Empirical analyses reveal strong cross-language structural regularities—demonstrating that syntactic graphs from languages as diverse as Python, Java, and Go can be aligned under a shared schema. We release the dataset publicly on Hugging Face\footnote{\url{https://huggingface.co/datasets/jugalgajjar/MultiLang-Code-Parser-Dataset}} and the accompanying codebase on GitHub\footnote{\url{https://github.com/JugalGajjar/MultiLang-Code-Parser-Dataset}}, which includes complete pipelines for dataset reproduction, grammar compilation, and a visualization tool for exploring the unified AST across languages. Together, these resources establish MLCPD as an open, reproducible foundation for future research in cross-language representation learning and program analysis.
\end{abstract}

\section{Introduction}

Understanding and reasoning over program structure across languages remains a central challenge in software intelligence. Despite recent advances in multilingual models such as CodeBERT~\cite{feng2020codebert} and GraphCodeBERT~\cite{guo2020graphcodebert}, there exists a persistent disconnect between language-agnostic modeling goals and the lack of structurally consistent, large-scale datasets. Most existing corpora—such as The Stack~\cite{stack}, StarCoder~\cite{starcoder}, and CodeSearchNet~\cite{codesearch}—focus on token-level or natural language alignments, overlooking fine-grained syntactic nodes and semantic categories that form the foundation for symbolic reasoning, program translation, and vulnerability detection.

The MLCPD dataset was created to address this gap by introducing a unified structural representation of code across ten major programming languages: \emph{C, C++, C\#, Go, Java, JavaScript, Python, Ruby, Scala, and TypeScript}. By integrating over seven million code files, MLCPD bridges syntax and semantics under a universal Abstract Syntax Tree (AST) schema. Each file is represented as a hierarchical JSON object preserving all syntactic relations, augmented with node-type statistics and semantic mappings for cross-language comparability. This uniform structure enables researchers to conduct language-agnostic analyses, trace syntactic parallels, and train models that reason across paradigms with minimal preprocessing overhead.

Our motivation stems from three key observations:  
(1) Syntax is a universal interface for reasoning about code, regardless of language surface form.  
(2) Modern LLM-based code models still lack interpretable and structurally aligned supervision signals.  
(3) A unified dataset that preserves language-specific nuances while enforcing schema normalization can empower research in cross-language program understanding, translation, and vulnerability analysis.

Beyond serving as a benchmark corpus, MLCPD aims to redefine how multilingual source code is represented and analyzed. Its universal schema provides a bridge between symbolic program representations and neural models, facilitating hybrid reasoning that leverages both structure and semantics. The dataset’s alignment across ten languages offers an unprecedented opportunity to study structural transfer, syntactic entropy, and multilingual generalization—laying the groundwork for the next generation of explainable, cross-language software intelligence systems.

\section{Related Work}

Research in multilingual program analysis has evolved considerably over the past two decades, transitioning from rule-based static analyzers to large-scale structural and representation-learning frameworks. Early efforts, such as cross-language refactoring systems~\cite{mayer2012cross} and dependency visualization tools~\cite{linos2003tool}, primarily targeted symbol extraction, interface linkage, and modular comprehension within multi-language codebases. These early systems provided valuable insight into heterogeneous software but were restricted to limited language pairs and lacked extensibility across modern ecosystems.

Subsequent frameworks, most notably LiSA~\cite{negrini2023lisa} and LARA~\cite{bispo2021multi}, introduced language-independent static analyzers by defining common intermediate representations (IRs) that could capture shared semantic structures. Universal intermediate representations (IRs) such as LLVM IR\cite{LLVM} and Java bytecode\cite{javabyte} provide language-agnostic abstractions at the compilation level. While these approaches advanced portability, they often relied on handcrafted grammars and deterministic rules, resulting in brittle scalability when applied to large, diverse corpora. Later developments such as AXA~\cite{roth2024axa} and PolyCruise~\cite{li2022polycruise} extended this idea to cross-language static and dynamic analysis by integrating multiple monolingual analyzers into unified pipelines. However, these frameworks remain tool-specific and execution-oriented rather than dataset-centric, offering limited value for statistical learning or code representation modeling.

In parallel, the introduction of permissively licensed corpora such as The Stack~\cite{stack} and StarCoder~\cite{starcoder} catalyzed the growth of language models for code generation, retrieval, and summarization. Despite their scale—covering terabytes of multilingual source code—these resources emphasize token-level data and metadata rather than syntactic or structural representations \cite{heh2024}. Consequently, while they have enabled progress in large language model (LLM) training, they provide little insight into structural patterns or cross-language regularities \cite{maw2023}. Smaller datasets like CodeSearchNet~\cite{codesearch} and Py150~\cite{py150} introduced limited AST supervision but remain narrow in scope, typically covering only one or two languages with incomplete structural annotations.

At the representation-learning level, recent models such as CodeBERT~\cite{feng2020codebert}, GraphCodeBERT~\cite{guo2020graphcodebert}, and AST-T5~\cite{ASTT5} have demonstrated the utility of incorporating structural signals through token–graph hybrids or attention over AST nodes. Yet, these models depend on heterogeneous parsing resources, each introducing language-specific biases and inconsistent node semantics. Without a standardized schema, embeddings learned from one language cannot be reliably transferred to another. This fragmentation reflects a core limitation across modern code intelligence research: the absence of a unified, lossless, and cross-language structural foundation.

The MultiLang Code Parser Dataset (MLCPD) directly addresses this gap. Unlike prior corpora that focus on textual data or tool-oriented analysis frameworks, MLCPD introduces a scalable, dataset-driven approach to structural unification. It integrates over seven million files across ten languages, each represented under a single, formally defined \emph{universal Abstract Syntax Tree (AST)} schema that preserves syntactic fidelity while enabling cross-language comparability. By decoupling parsing from modeling and by providing standardized, schema-validated structures, MLCPD lays the groundwork for reproducible cross-language learning and fine-grained structural reasoning at scale. Table~\ref{tab:related-comparison} summarizes how MLCPD contrasts with prior categories of research, highlighting its combination of universality, losslessness, and scalability.

\begin{table}[H]
\centering
\caption{Comparison of prior work categories w.r.t. MLCPD goals ($\checkmark$: present, $\times$: absent).}
\label{tab:related-comparison}
\vspace{0.3em}
\setlength{\tabcolsep}{6pt}
\renewcommand{\arraystretch}{1.1}
\begin{tabular}{lcccc}
\toprule
\textbf{Category} & \textbf{Structural Rep.} & \textbf{Cross-Lang.} & \textbf{Lossless} & \textbf{Uniformity} \\
\midrule
Code Datasets (The Stack, etc.) & $\times$ & $\times$ & $\times$ & $\times$ \\
Parsing Frameworks (Tree-sitter) & $\checkmark$ & $\times$ & $\checkmark$ & $\times$ \\
IRs (LLVM) & $\checkmark$ & $\checkmark$ & $\times$ & $\times$ \\
\textbf{MLCPD} & $\checkmark$ & $\checkmark$ & $\checkmark$ & $\checkmark$ \\
\bottomrule
\end{tabular}
\end{table}

\section{Dataset Design and Implementation}
\label{sec:dataset_design_implementation}

The design and construction of the MultiLang Code Parser Dataset (MLCPD) was guided by a single unifying principle: to establish a \emph{universal, lossless, and language-agnostic representation} of program structure that enables consistent analysis across ten programming languages—C, C++, C\#, Go, Java, JavaScript, Python, Ruby, Scala, and TypeScript. Achieving this required not only scalable data collection, but also the creation of a new abstraction layer over existing parsing technologies that could reconcile their inherent syntactic diversity.

\subsection{Data Collection and Preprocessing}
MLCPD builds upon permissively licensed source files from the StarCoder dataset~\cite{starcoder}, filtered under the MIT, Apache-2.0, and BSD licenses. Each file underwent multi-stage preprocessing: (1) normalization of character encoding to UTF-8, (2) removal of Byte Order Marks, non-printable characters, and redundant whitespace, and (3) statistical filtering to exclude trivial and auto-generated code. Using an interquartile-range (IQR) filter, we eliminated outliers and retained only files containing between 10 and 10,000 lines, ensuring a corpus of non-trivial, parseable, and representative programs.  

Following preprocessing, all code samples were revalidated for syntactic completeness and deduplicated via content-based hashing, ensuring that each source file corresponds to a unique structural representation. This normalization process established a high-quality foundation for large-scale parsing and schema inference.

\subsection{Design Philosophy and Universal Schema}
Our design goal extended far beyond simple code parsing—it sought to encode the \emph{entire syntactic structure} of diverse languages in a shared representational space. To achieve this, we defined four non-negotiable design criteria:  
\begin{itemize}
    \item \textbf{Losslessness:} Preserve every syntactic element (tokens, delimiters, punctuation, whitespace) without semantic compression.  
    \item \textbf{Uniformity:} Enforce a consistent JSON schema across all languages.  
    \item \textbf{Queryability:} Enable efficient and language-independent structural querying.  
    \item \textbf{Scalability:} Support millions of files with minimal memory overhead.  
\end{itemize}

Prior to implementation, extensive empirical analysis was conducted over 5,000 sample files (500 per language) to identify structural commonalities. While earlier approaches to unification attempted to impose semantic alignment (e.g., mapping class inheritance across paradigms), we found that such methods violated the losslessness principle. Instead, MLCPD’s innovation lies in enforcing \emph{structural homogeneity without suppressing syntactic heterogeneity}. This insight led to the design of a four-layer universal AST schema.

\paragraph{Layer 1: Metadata Block.}  
Captures global file characteristics and parser integrity diagnostics, serving both analytical and validation roles. \\

\begin{lstlisting}[caption={Example metadata block for a parsed source file.}]
"metadata": {
  "lines": 247,
  "avg_line_length": 34.2,
  "nodes": 1853,
  "errors": 0,
  "source_hash": "a3f5e8c9..."
}
\end{lstlisting}

Each field quantifies core properties: line count and average length indicate code density, node count measures syntactic complexity, and errors reflect parser resilience. The \texttt{source\_hash} field uses SHA-256 for deterministic fingerprinting, chosen over weaker alternatives (MD5, SHA-1) for its collision resistance~\cite{nist2015, velvindronl2021}. This facilitates reproducible node-level identification, content-addressable deduplication, and incremental dataset updates.

\paragraph{Layer 2: Flat Node Array.}  
Maps the entirety of the file's syntactic structure into a list of atomic, directly-addressable nodes. This process of linearizing the tree structure significantly boosts the efficiency of code analysis and traversal operations.

\begin{lstlisting}[caption={Representative AST node structure.}]
"nodes": [
  {
    "id": 0,
    "type": "function_definition",
    "text": "def calculate_sum(a, b):\n    return a + b",
    "parent": null,
    "children": [1, 5, 6],
    "start_point": {"row": 10, "column": 0},
    "end_point": {"row": 11, "column": 20}
  }
]
\end{lstlisting}

This representation allows constant-time ($O(1)$) traversal between nodes, supports vectorized analysis enabling parallel processing, and ensures serialization stability guaranteeing consistent representation across processing environments.

\paragraph{Layer 3: Node Categorization.}  
Groups syntactic nodes into a universal three-level taxonomy—\texttt{declarations}, \texttt{statements}, and \texttt{expressions}—representing the highest shared abstraction among all ten languages.

\begin{lstlisting}[caption={Categorical indexing of declarations, statements, and expressions.}]
"node_categories": {
  "declarations": {"functions": [0, 42], "classes": [120, 256]},
  "statements": {"loops": [65, 78], "returns": [35, 140]},
  "expressions": {"calls": [25, 36], "identifiers": [10, 11]}
}
\end{lstlisting}

This enables direct category-based retrieval (e.g., all function definitions) without recursive traversal, which is critical for large-scale analytical queries.

\paragraph{Layer 4: Cross-Language Map.}  
Abstracts language-specific constructs into universal schema roles while maintaining the original syntax.

\begin{lstlisting}[caption={Cross-language normalization of function and class declarations.}]
"cross_language_map": {
  "function_declarations": [
    {"node_id": 0, "universal_type": "function", "name": "calculate_sum"}
  ],
  "class_declarations": [
    {"node_id": 120, "universal_type": "class", "name": "DataProcessor"}
  ]
}
\end{lstlisting}

This layer bridges syntactic differences, enabling structurally equivalent constructs across languages (e.g., Python’s \texttt{def} and Java’s \texttt{public static void}) to be queried uniformly.

\begin{figure}[H]
\centering
\includegraphics[width=\linewidth]{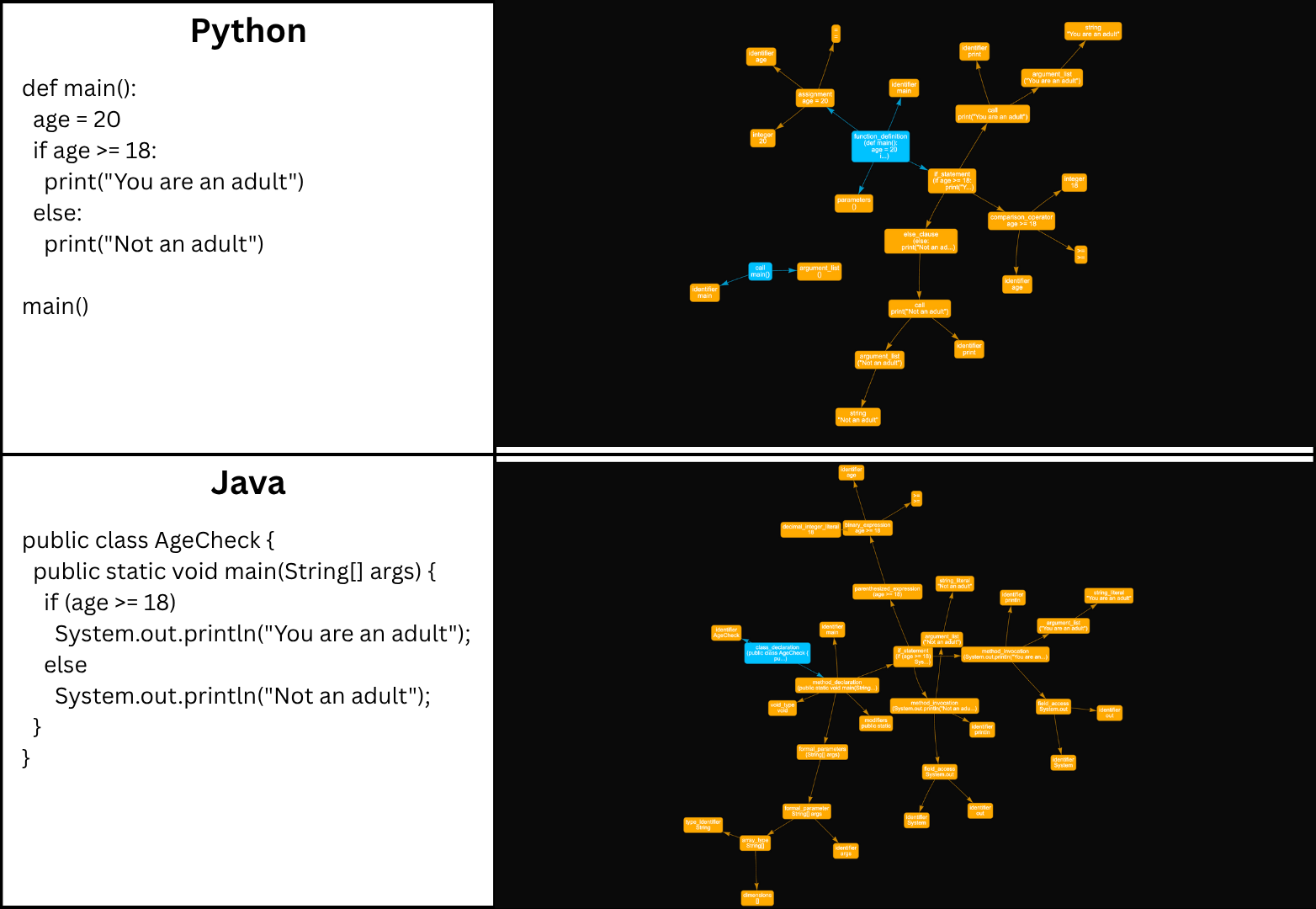}
\caption{Visualization of Python and Java ``Age Check'' programs parsed under the universal schema, demonstrating consistent alignment across languages.}
\label{fig:python-java-agecheck}
\end{figure}

\subsection{Empirical Schema Validation and Parser Evolution}
Initial prototypes used rule-based language mappers and regular-expression parsers to define the schema mapping. While these provided limited coverage, both approaches proved brittle and lossy. Rule-based mappers captured only predefined patterns and required quadratic maintenance as languages increased, while regex-based approaches failed on nested constructs and omitted non-captured syntax.  

The breakthrough came with the adoption of Tree-sitter~\cite{tree-sitter}, an incremental parsing system with hand-crafted, formally verified grammars. Tree-sitter offered three indispensable features: (1) complete syntactic coverage, (2) a uniform parsing interface across languages, and (3) robust error recovery for real-world code fragments. These properties satisfied all four schema design principles, as shown in Table~\ref{tab:parsing-comparison}.

\begin{table}[H]
\centering
\caption{Comparison of parsing techniques with respect to schema design requirements.}
\label{tab:parsing-comparison}
\begin{tabular}{lcccc}
\toprule
\textbf{Technique} & \textbf{Lossless} & \textbf{Uniform} & \textbf{Queryable} & \textbf{Scalable} \\
\midrule
Rule-Based Mappers & $\times$ & $\times$ & $\times$ & $\times$ \\
Regex-Based Parsing & $\times$ & $\checkmark$ & $\checkmark$ & $\checkmark$ \\
Tree-sitter Grammars & $\checkmark$ & $\checkmark$ & $\checkmark$ & $\checkmark$ \\
\bottomrule
\end{tabular}
\end{table}

\subsection{Parsing and Normalization Pipeline}
All ten Tree-sitter grammars were compiled into a shared dynamic library, ensuring uniform parsing behavior. The processing pipeline comprises six deterministic stages:  
(1) language detection and grammar dispatch,  
(2) recursive AST extraction,  
(3) node categorization,  
(4) cross-language mapping,  
(5) JSON schema validation, and  
(6) Parquet serialization for scalable storage.  

Algorithm~\ref{alg:extract_ast_structure} and Algorithm~\ref{alg:create_cross_language_map} summarize the extraction and mapping stages in Tree-sitter–based parsing.

\begin{algorithm}[H]
\caption{Extract AST Structure}
\label{alg:extract_ast_structure}
\DontPrintSemicolon
\KwIn{Tree $T$, source string $S$, language $L$}
\KwOut{Hierarchical list of AST nodes}
Initialize empty list $N$ and counter $id \leftarrow 0$\;
\Fn{\textsc{TraverseNode}($node, parent$)}{
    \If{ShouldSkipNode($node$)}{
        \ForEach{$child$ in $node.children$}{
            \textsc{TraverseNode}($child, parent$)\tcp*{Skip delimiters}
        }
        \Return{}
    }
    Create dictionary $d$ with fields:\;
    $d[\texttt{id}] \gets id$; $d[\texttt{type}] \gets node.type$; $d[\texttt{text}] \gets S[node.start\_byte : node.end\_byte]$;\;
    $d[\texttt{parent}] \gets parent$; $d[\texttt{children}] \gets [\,]$\;
    \ForEach{$child$ in $node.children$}{
        $cid \gets$ \textsc{TraverseNode}($child, id$)\;
        \If{$cid \neq$ None}{append $cid$ to $d[\texttt{children}]$}
    }
    append $d$ to $N$; $id \gets id + 1$\;
    \Return $d[\texttt{id}]$\;
}
\textsc{TraverseNode}($T.root$, None)\;
\Return $\{\,\texttt{"nodes"}: N\,\}$\;
\end{algorithm}

\begin{algorithm}[H]
\caption{Create Cross-Language Map}
\label{alg:create_cross_language_map}
\DontPrintSemicolon
\KwIn{AST structure $A$, categories $C$, language $L$}
\KwOut{Cross-language map $\mathcal{M}$}
Initialize $\mathcal{M} \gets \{\texttt{"function\_declarations"} : [\,]\}$\;
\ForEach{$f$ in $C[\texttt{declarations}][\texttt{functions}]$}{
    $n \gets$ GetNodeById($f$); $name \gets$ ExtractNameFromText($n[\texttt{text}], L$)\;
    Append $\{\texttt{"node\_id"}:f, \texttt{"universal\_type"}:"function", \texttt{"name"}:name, \texttt{"text\_snippet"}:n[\texttt{text}][:100]\}$ to $\mathcal{M}[\texttt{"function\_declarations"}]$\;
}
\Return $\mathcal{M}$\;
\end{algorithm}

\subsection{Dataset Statistics and Structural Insights}
After normalization, MLCPD comprises over seven million records (Table~\ref{tab:dataset-overview}), distributed evenly across ten languages. The dataset occupies approximately 114 GB in Parquet format and 600 GB in-memory—an indication of the balance between structural richness and compression efficiency.

\begin{table}[H]
\centering
\caption{Dataset overview.}
\label{tab:dataset-overview}
\begin{tabular}{lc}
\toprule
\textbf{Metric} & \textbf{Value} \\
\midrule
Total Languages & 10 \\
Total Files & 40 \\
Total Records & 7,021,722 \\
Disk Size & $\sim$114 GB (Parquet) \\
Memory Size & $\sim$600 GB (in-memory) \\
\bottomrule
\end{tabular}
\end{table}

\begin{figure}[H]
\centering
\includegraphics[width=0.55\linewidth]{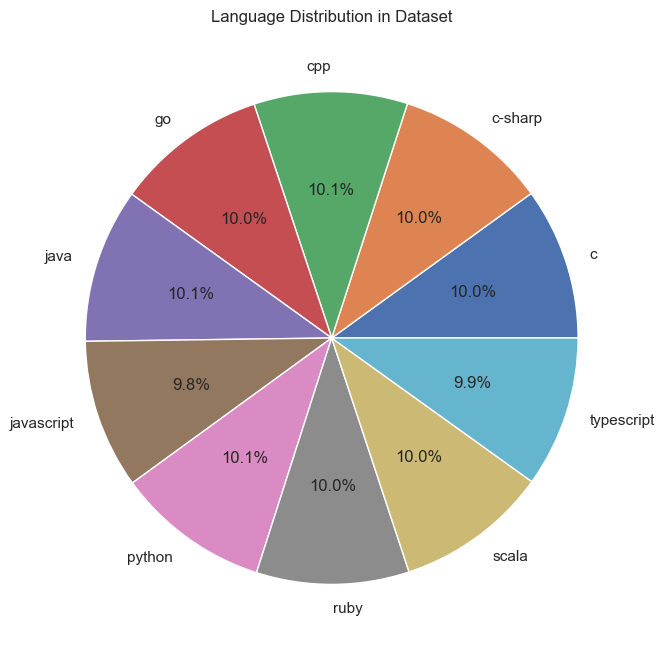}
\caption{Distribution of records across ten programming languages. The uniform proportions validate balanced representation.}
\label{fig:pie-language}
\end{figure}

Parsing reliability remains exceptional, with 7,021,718 out of 7,021,722 files parsed successfully (99.99994\%), as summarized in Table~\ref{tab:conversion-stats}. This near-perfect conversion rate underscores the robustness of the pipeline.

\begin{table}[H]
\centering
\caption{Conversion reliability statistics.}
\label{tab:conversion-stats}
\begin{tabular}{l l}
\toprule
Successful conversions & 7,021,718 \\
Total attempted & 7,021,722 \\
Failures & 4 \\
Success rate & 99.99994\% \\
\bottomrule
\end{tabular}
\end{table}

\begin{figure}[H]
\centering
\includegraphics[width=0.6\linewidth]{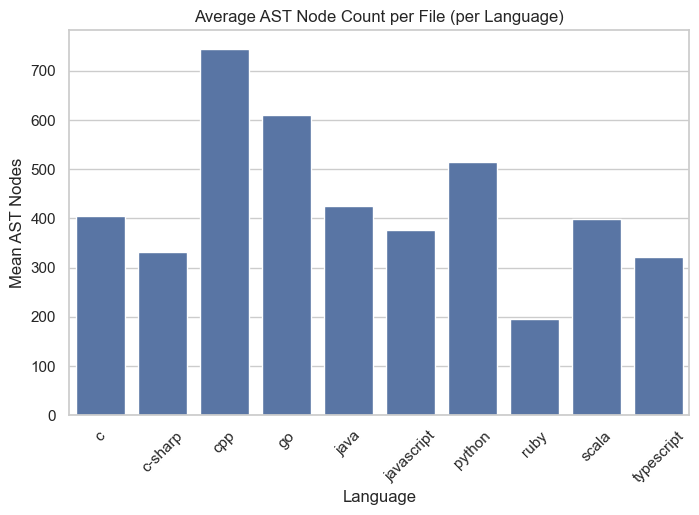}
\caption{Average AST node count per file across languages. Higher counts reflect verbose syntactic forms and explicit typing.}
\label{fig:avg-nodes}
\end{figure}

Figure~\ref{fig:avg-nodes} compares the average AST node count per file, revealing structural verbosity patterns—C++ and Go produce denser syntactic trees than Python or Ruby due to explicit type and control structures.  

\begin{figure}[H]
\centering
\includegraphics[width=0.6\linewidth]{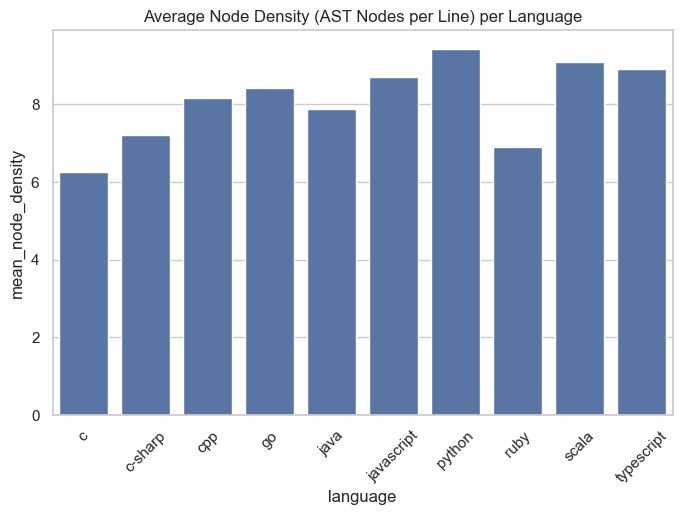}
\caption{Average node density per language, reflecting syntactic compactness and complexity.}
\label{fig:avg-density}
\end{figure}

Node density (nodes per line) in Figure~\ref{fig:avg-density} reflects each language’s syntactic granularity. Python and Scala show higher densities due to indentation-based structuring and implicit typing, which decompose compact expressions into multiple AST nodes. In contrast, Go and Java achieve lower densities through explicit type declarations and block delimiters, indicating that the schema faithfully captures linguistic verbosity without distortion.

\begin{figure}[H]
\centering
\includegraphics[width=0.75\linewidth]{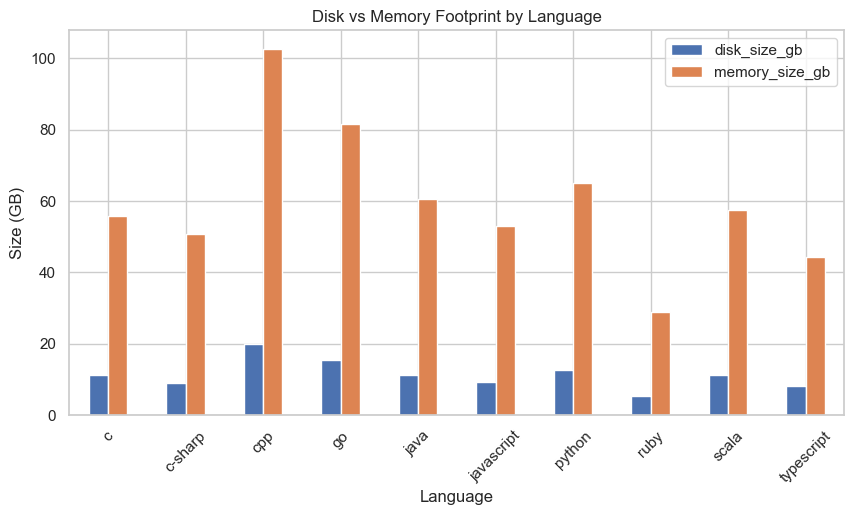}
\caption{Disk versus memory footprint per language, confirming consistent compression ratios across corpora.}
\label{fig:disk-memory}
\end{figure}

Figure~\ref{fig:disk-memory} compares disk and memory footprints, revealing a stable compression ratio of roughly $5.5\times$ across all languages. This consistency indicates that the universal schema introduces uniform structural overhead regardless of syntax complexity. The result confirms MLCPD’s scalability and demonstrates that schema normalization yields balanced, storage-efficient representations across diverse programming paradigms.

\subsection{Cross-Language Structural Correlation}
\begin{figure}[H]
\centering
\includegraphics[width=0.8\linewidth]{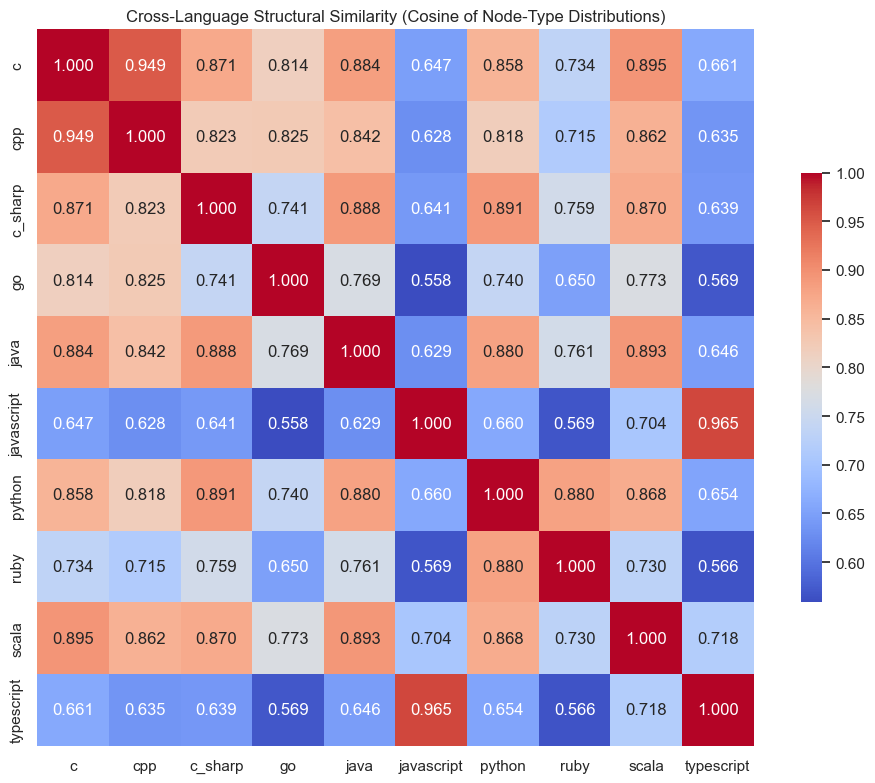}
\caption{Cross-language structural similarity (cosine of node-type distributions).}
\label{fig:cross-lang-heatmap}
\end{figure}

The cosine similarity matrix in Figure~\ref{fig:cross-lang-heatmap} quantifies how closely languages align in their syntactic structure under the universal schema. High similarity scores ($>$0.90) between C and C++ indicate that these statically typed, statement-driven languages share closely matching structural patterns, driven by explicit type declarations and block-based syntax. C\# and Scala also exhibit strong alignment with Java (0.87–0.89), reflecting their object-oriented and strongly typed design lineage.  
In contrast, JavaScript and TypeScript form a distinct high-similarity cluster ($0.96$), confirming their near-identical grammatical roots within prototype-based programming. Python and Ruby demonstrate moderate mutual similarity ($0.88$) but diverge structurally from compiled languages due to their expression-oriented, dynamically typed nature. Go, while syntactically simpler, remains closer to C-family languages (0.76–0.82), supporting its design heritage. Overall, these correlations validate that MLCPD’s universal schema captures genuine cross-language syntactic relationships rather than superficial token overlaps.

\begin{figure}[H]
\centering
\includegraphics[width=0.6\linewidth]{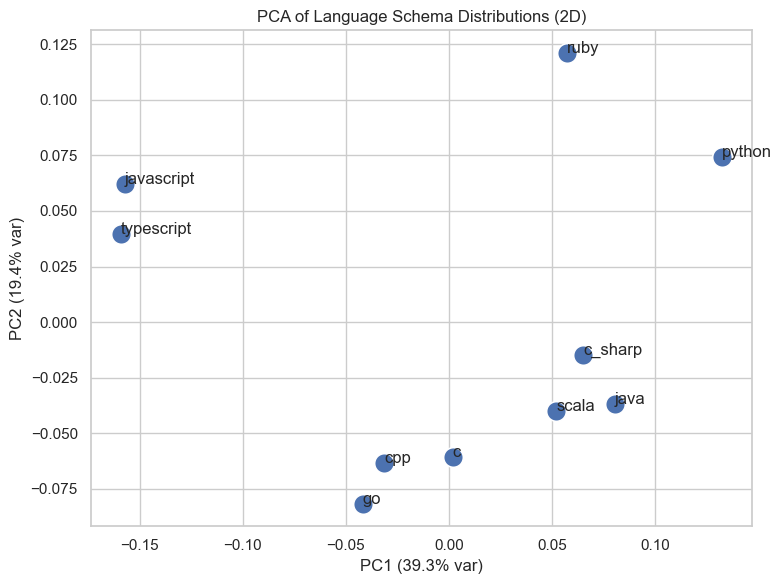}
\caption{PCA projection of language schema embeddings, revealing natural syntactic clusters.}
\label{fig:pca-schema}
\end{figure}

Finally, principal component projection (Figure~\ref{fig:pca-schema}) visualizes schema embeddings, where semantically proximate languages (e.g., Java and Scala; JavaScript and TypeScript) occupy neighboring regions in latent space—strong evidence that MLCPD captures deep syntactic link across languages.

\section{Qualitative Example: Python vs. Java “Age Check” Schema}

To demonstrate the universality of MLCPD’s schema,
we visualize equivalent Python and Java programs that determine if a person is an adult.
Despite lexical and syntactic differences, their underlying ASTs align seamlessly under the same universal schema (shown in Figure \ref{fig:python-java-agecheck}).

Both code samples produce function nodes with identical hierarchical roles:
parameter declarations, conditional statements, and return expressions.
The cross-language map unifies these constructs through shared \texttt{"function"}, \texttt{"conditional"}, and \texttt{"literal"} node types, validating schema coherence.

\section{Discussion and Limitations}

MLCPD is designed as a research-grade corpus rather than an exhaustive code archive.
While it ensures near-perfect conversion rates and schema uniformity, some limitations remain:
\textit{(a)} language coverage is currently limited to ten high-impact languages;
\textit{(b)} semantic annotations (e.g., data-flow, control-flow edges) are not yet included;
and \textit{(c)} minor discrepancies in indentation and tokenization may persist across certain Tree-sitter grammars.

Nonetheless, the dataset establishes a foundational layer for building
cross-language representation models, interpretable code embeddings, and multilingual static analysis tools.

\section{Conclusion}

We have presented MLCPD, a large-scale, structurally uniform dataset for multilingual code understanding. By constructing a universal AST schema that spans ten programming languages, MLCPD bridges the divide between symbolic program structure and data-driven modeling. Its high conversion accuracy, balanced coverage, and visualized structural correspondences make it a key resource for future research in multilingual program analysis, cross-language translation, and hybrid graph-language modeling. The dataset and scripts are open-sourced on \href{https://huggingface.co/datasets/jugalgajjar/MultiLang-Code-Parser-Dataset}{Hugging Face} and \href{https://github.com/JugalGajjar/MultiLang-Code-Parser-Dataset}{GitHub} for community use.

\end{document}